\documentclass{rmaa}

\title{Radio Proper Motions of Wolf-Rayet Stars}

 \author{ Sergio Dzib\altaffilmark{1} and Luis F. Rodr\'\i guez\altaffilmark{1}}
  \altaffiltext{1}{Centro de Radioastronom\'{\i}a y Astrof\'{\i}sica, Universidad
Nacional Aut\'onoma de M\'exico, Campus Morelia} 

\fulladdresses{
\item Sergio Dzib and Luis F. Rodr\'\i guez: Centro de Radiostronom\'{\i}a
y Astrof\'\i sica, 
UNAM, A. P. 3-72, 
(Xangari), 58089 Morelia, Michoac\'an, M\'exico
(l.rodriguez; sdzib@astrosmo.unam.mx).
}

\shortauthor{Dzib \& Rodr\'\i guez}
\shorttitle{Proper Motions of Wolf-Rayet Stars}

\SetVolume{45} \SetFirstPage{01} \SetYear{2009}
\ReceivedDate{\today} 
\AcceptedDate{Year Month Day} 

\resumen{Presentamos el an\'alisis de observaciones tomadas del archivo del Very Large
Array de seis estrellas Wolf-Rayet con emision de radio, con el prop\'osito de 
determinar sus movimientos propios. T\'\i picamente, estas observaciones cubren
periodos de tiempo del orden de 10 a 20 a\~nos. Para verificar el m\'etodo,
incluimos a WR 140 en la muestra, encontrando que los
movimientos propios determinados por nosotros son
varias veces m\'as precisos que los determinados por
Hipparcos y consistentes con ellos dentro del ruido observacional.
Las otras cinco estrellas WR no fueron estudiadas por Hipparcos y reportamos
sus movimientos propios por primera vez. Los movimientos propios de
WR 145a = Cyg X-3 son consistentes con que la fuente est\'e estacionaria
en su marco de reposo local y sugieren que el hoyo negro en este sistema
binario se form\'o por colapso directo de una estrella masiva, sin la expulsi\'on de una
remanente de supernova.}
 
\abstract{We present the analysis of observations taken from the
Very Large Array archive of six Wolf-Rayet stars with radio emission,
with the purpose of determining their proper motions.
Typically, these observations cover periods of 10 to 20 years.
To verify the method, we included WR 140 in the sample, finding that
the proper motions determined by us are
a few times more accurate than and
consistent within noise with those of Hipparcos.
The other five WR stars were not studied by Hipparcos and we report their
proper motions for the first time.
The proper motions for WR 145a = Cyg X-3 are consistent with the source
being stationary with respect to its local standard of rest and suggest that
the black hole in this binary system formed by direct collapse
of a massive star, without expulsion of a supernova remnant.
}
      

\begin{document}

\maketitle

\section{Introduction}

The Hipparcos satellite obtained accurate proper motions of 118,218 
stars with visual magnitudes brighter than 12.5. Fainter stars
were not studied by this mission and in many cases these objects do not
have measured proper motions from previous studies. 
Some objects that are faint in the visible (for example, by obscuration or
large distance) are easy to detect at radio wavelengths and proper motions
can be obtained accurately for the first time from the analysis of radio data.

In this paper we present the analysis of six Wolf-Rayet stars with radio continuum
emission, that have been observed with high angular resolution 
at the Very Large Array (VLA) in several occasions over the years, with the purpose of
determining their proper motions.

The star WR 140 was included in our study, even when it has
an accurate Hipparcos proper motion determination 
(Perryman et al. 1997), to check the reliability of
the technique used. The other five stars have visual magnitudes larger than 
12.5 and their proper motions are not reported by Hipparcos.
Besides determining their proper motions for the first time, we also had
an interest in finding additional examples of the relatively
small class of massive runaway stars,
such as those discussed by Moffat et al. (1998).

\section{Observations}

We searched in the archives of the Very Large Array (VLA) of the 
NRAO\footnote{The National Radio Astronomy Observatory is operated
by Associated Universities Inc. under cooperative agreement with
the National Science Foundation.} for observations of WR stars
taken with high angular resolution (0\rlap.{''}4 or better). 
For most of the observation we used the A configuration of the 
VLA, except for WR 140 in the epochs 1990.43 and 1999.76, and for 
WR 112 in the epoch 2001.14. In these cases the configuration was BnA.


For each source, we needed that the same phase calibrator
was used in all epochs to obtain
accurate absolute astrometry. These phase calibrators are 
listed in Table 1, with the latest refined positions from the VLA Calibrator
Manual adopted
in the reduction of all the epochs. 
For each source presented here, we found at least three epochs
with these requirements.

The data were calibrated using the standard routines within the Astronomical Image Processing 
System (AIPS). Systematic errors in the order of 5 to 20 mas were added in quadrature
to the formal errors of the fitting task JMFIT to obtain reduced $\chi^2$ values of
1 in the linear least squares fits to the proper motions.

\begin{table*}[htbp]
\small
  \setlength{\tabnotewidth}{1.0\columnwidth} 
  \tablecols{5} 
  \caption{Phase calibrators}
  \begin{center}
    \begin{tabular}{lcccc}\hline\hline
             &\multicolumn{2}{c}{Adopted Position} & Source & Angular  \\
\cline{2-3}
Calibrator     &$\alpha(2000)$ & $\delta(2000)$ & Observed       & Distance ($^\circ$)$^a$ \\
\hline\hline
2007+404 & $20^h07^m44\rlap.{^s}945$ & $+40^\circ 29'48\rlap.{''}604$  & WR140  & 4.8 \\ 
2007+404 & $20^h07^m44\rlap.{^s}945$ & $+40^\circ 29'48\rlap.{''}604$  & WR145a & 6.7 \\ 
2007+404 & $20^h07^m44\rlap.{^s}945$ & $+40^\circ 29'48\rlap.{''}604$  & WR146  & 7.5 \\ 
2007+404 & $20^h07^m44\rlap.{^s}945$ & $+40^\circ 29'48\rlap.{''}604$  & WR147  & 7.5 \\ 
1925+211 & $19^h25^m59\rlap.{^s}605$ & $+21^\circ 06'26\rlap.{''}162$  & WR125  & 1.7 \\
1820-254 & $18^h20^m57\rlap.{^s}849$ & $-25^\circ 28'12\rlap.{''}584$  & WR112  & 6.6 \\
\hline\hline
\tabnotetext{a}{Angular distance between the phase calibrator and the  source.}
    \label{tab:pc}
    \end{tabular}
  \end{center}
\end{table*}

\section{Results}

The equatorial proper motions determined by us for the six WR stars are given in Table 2.
In this Table we also give the 
distances to the stars, taken from van der Hucht (2001), with the
exception of WR~140 where the distance was taken from Dougherty et al.
(2005).

In Table 3, we give the observed proper motions in galactic coordinates,
the expected galactic proper motions, and the difference between
the observed and the expected galactic proper motions.
To obtain the expected galactic
proper motions we used
the galactic rotation model of Brand \& Blitz (1993) and the
velocity of the Sun with respect to the local standard of rest from Dehnen \& Binney (1998).

A comparison between the measured galactic proper motions and the expected galactic
proper motions (see columns 6 and 7 of Table 3)
indicates that four (WR 125, WR 145a, WR 146, and WR 147) of the six stars 
have observed galactic proper motions consistent with the stars being approximately
stationary with respect to their respective local standard of rest.
In the case of WR~140 and WR~112, we obtain significant differences between
the expected and the observed galactic
proper motions, as discussed below.
We have used as a criterion for a runaway star to depart by more than $42~km~s^{-1}$ from its
expected local standard of rest velocity (Moffat et al. 1998).

All six systems studied are binaries. One can then ask if the radio emission is
originating from one or both stars, or even from the interacting wind shock region.
In the case of WR~145a, the binary separation is of order $10^{11}$ cm
(Achterberg 1989), equivalent
to $\sim$0.001 mas at a distance of 9 kpc. This angular separation is much smaller than our
typical positional errors ($\sim$10-20 mas) and we can consider this binary as a 
point source. Even in the case of WR~140, WR~112, WR~125, and WR~146, where the
binary separation is expected to be of order $4 \times 10^{14}$ cm,
the angular separations are expected to be, for distances of order 1 kpc
or larger, smaller than $\sim$30 mas. This is comparable with our positional
errors, but much smaller than our angular resolution of $\sim$300 mas.
We are then unable to correct for this effect and assume it will not significantly
alter our results. Finally, in the case of WR~147, the angular separation
is large enough to allow separate imaging of the southern star and of the
northern shock interaction zone, and we can perform individual astrometry for both components.

In the next section we discuss each source individually.

\begin{table*}
  \setlength{\tabnotewidth}{1.0\columnwidth} 
  \tablecols{6} 
     \caption{Equatorial proper motions of the WR stars studied}
  \begin{center}
    \begin{tabular}{lccccc}
\hline\hline
        &  $l$ & $b$   &   d   &   $cos(\delta) \mu_{\alpha}$    
&  $\mu_{\delta}$ \\
Star  &   $(^\circ)$    &  $(^\circ)$  & (kpc) & (mas yr$^{-1}$) & (mas yr$^{-1}$) \\

\hline\hline
WR 112  & 12.146  & -1.186  &  4.15 & -11.2$\pm$3.1  & -13.5$\pm$5.5 \\
WR 125  & 54.445  & 1.058   &  3.06 & -2.7$\pm$0.5  & -5.7$\pm$0.6 \\
WR 140  & 80.930  & 4.177   & 1.85$^a$ & -5.0$\pm$0.2  & -1.2$\pm$0.1 \\
WR 145a & 79.846  & 0.700 & 9.00 & -2.5$\pm$0.2  & -4.3$\pm$0.3 \\
WR 146  & 80.561  & 0.445  & 0.72 & -5.2$\pm$0.3  & -2.5$\pm$2.1 \\
WR 147  & 79.848  & -0.315 &  0.65 & -2.0$\pm$0.6  & -5.1$\pm$1.0 \\
\hline\hline
\tabnotetext{a}{From Dougherty et al. 2005.}
    \end{tabular}
   \end{center}
\end{table*}

\begin{table*}
  \setlength{\tabnotewidth}{2.0\columnwidth} 
  \tablecols{7} 
     \caption{Galactic proper motions of the WR stars studied}
  \begin{center}
    \begin{tabular}{lcccccc}
\hline\hline
        &   $cos(b) \mu_l(o)^a$ & $\mu_b(o)^a$ &
$cos(b) \mu_l(e)^b $ & $\mu_b(e)^b$ & $\Delta [cos(b) \mu_l(o-e)]^c$ &
$\Delta [\mu_b(o-e)]^c$ \\
Star  &   (mas yr$^{-1}$)
& (mas yr$^{-1}$) & (mas yr$^{-1}$) & (mas yr$^{-1}$) & (mas yr$^{-1}$) & (mas yr$^{-1}$)      \\

\hline\hline
WR 112  &  -17.2$\pm$5.1  & +3.4$\pm$3.8 & -1.1 & -0.3 & -16.1$\pm$5.1 & +3.7$\pm$3.8 \\
WR 125  &  -6.3$\pm$0.6  & -0.3$\pm$0.5 & -4.5 & -0.5 & -1.8$\pm$0.6 & +0.2$\pm$0.5  \\
WR 140  &  -3.8$\pm$0.1  & +3.5$\pm$0.2 & -4.4 & -0.8 & +0.6$\pm$0.1 & +4.3$\pm$0.2 \\
WR 145a &  -4.9$\pm$0.3  & -0.5$\pm$0.2 & -4.2 & -0.2 & -0.7$\pm$0.3 & -0.3$\pm$0.2  \\
WR 146  &  -5.1$\pm$1.7  & +2.7$\pm$1.3 & -2.8 & -2.1 & -2.3$\pm$1.7 & +5.0$\pm$1.3 \\
WR 147  &  -5.3$\pm$0.9  & -1.5$\pm$0.8 & -2.5 & -2.3 & -2.8$\pm$0.9 & +1.3$\pm$0.8  \\
\hline\hline
\tabnotetext{a}{Observed galactic proper motions.}
\tabnotetext{b}{Expected galactic proper motions from model discussed in text.}
\tabnotetext{c}{Difference between the observed and the expected galactic proper motions.}
    \end{tabular}
   \end{center}
\end{table*}

\section{Comments on Individual Sources}

\subsection{WR 140}

This is a WR+O colliding-wind binary system with a 7.9
year period, that has been studied in detail
by Dougherty et al. (2005) with VLBA observations. These authors find 
a semimajor axis of $9.0 \pm 0.5$ mas for the orbit in the plane of the sky,
a total mass of $74 \pm11$ $M_\odot$, and a distance of $1.85 \pm 0.16$ kpc.   

In order to check the reliability of our method, we did a determination of the proper motions
of this star, that has previously reported optical and radio
determinations. We used the epochs given in Table 4. 
The positions of the star as a function of time are shown in Figure 1.
The proper motions of this star were first studied by the Hipparcos satellite
(Perryman et al. 1997), 
that obtained
$cos(\delta) \mu_{\alpha}=-5.26\pm0.58\ mas\ yr^{-1}$ and 
$\mu_{\delta}=-2.37\pm0.49\ mas\ yr^{-1}$. More recently Boboltz et al. (2007), 
used the VLA plus Pie Town, to measure its proper motion at radio frecuencies. 
They obtained $cos(\delta) \mu_{\alpha}=-4.72\pm0.66\ mas\ yr^{-1}$ and $\mu_{\delta}=-1.89\pm0.64\ mas\ yr^{-1}$.
We obtained $cos(\delta) \mu_{\alpha}=-5.0\pm0.2\ mas\ yr^{-1}$ and $\mu_{\delta}=-1.2\pm0.1\ mas\ yr^{-1}$.
Our determination is a factor of a few more accurate than the previous measurements
and coincides with them at the 1 to 2-$\sigma$ error level. 

Using VLBA observations, Dougherty et al. (2005) have determined accurate proper motions
for the wind-collision region, from where most of the radio emission originates.
Their values, using only observations obtained during the night,
give $cos(\delta) \mu_{\alpha}=-5.44\pm0.25\ mas\ yr^{-1}$ and $\mu_{\delta}=-0.84\pm0.46\ mas\ yr^{-1}$,
are very similar to those obtained by us. 

Comparison between the observed and the expected
galactic proper motions given in Table 3, indicates that WR 140
has a peculiar motion of orden $\Delta \mu_{b} \simeq +4.3\pm0.2\ mas\ yr^{-1}$, that at a
distance of 1.85 kpc is equivalent to a peculiar velocity of $\sim34\pm2~km~ s^{-1}$. 
Practically all of the peculiar velocity is in the direction of positive galactic latitude.
Since WR 140 is at a galactic latitude of $b = +4\rlap.^\circ2$, these results
suggest that it may be moving away from an original position located in the galactic plane, 
near $l = +80\rlap.^\circ9$. In any case, the peculiar velocity is below our 
adopted velocity criterion for a runaway star. 

\begin{table*}[htbp]
\small
  \setlength{\tabnotewidth}{1.0\columnwidth} 
  \tablecols{4} 
  \caption{Archive Data for WR 140}
  \begin{center}
    \begin{tabular}{lccc}\hline\hline
                      &         & Wavelength & Beam  \\
Epoch                 & Project & (cm)& Angular Size $^a$ \\ 
\hline
1983 Oct 30 (1983.83) & AB252	& 6.0	& $0\rlap.{''}36 \times 0\rlap.{''}33;~\ 19^\circ$  \\
1990 Jun 07 (1990.43) & VH54G 	& 6.0 & $0\rlap.{''}47 \times 0\rlap.{''}36;~\ 68^\circ$  \\
1999 Oct 03 (1999.76) & BB117 	& 3.6 & $0\rlap.{''}31 \times 0\rlap.{''}24;~\ 79^\circ$  \\
2004 Oct 18 (2004.80) & AJ315 	& 3.6 & $0\rlap.{''}26 \times 0\rlap.{''}11;~\ 71^\circ$  \\
2006 Apr 22 (2006.31) & BD114 	& 3.6 & $0\rlap.{''}22 \times 0\rlap.{''}17;~\ -7^\circ$  \\
\hline\hline
\tabnotetext{a}{Major axis $\times$ minor axis; position angle.}
    \label{tab:1}
    \end{tabular}
  \end{center}
\end{table*}

\begin{figure*}
\centering
\includegraphics[scale=0.4, angle=0]{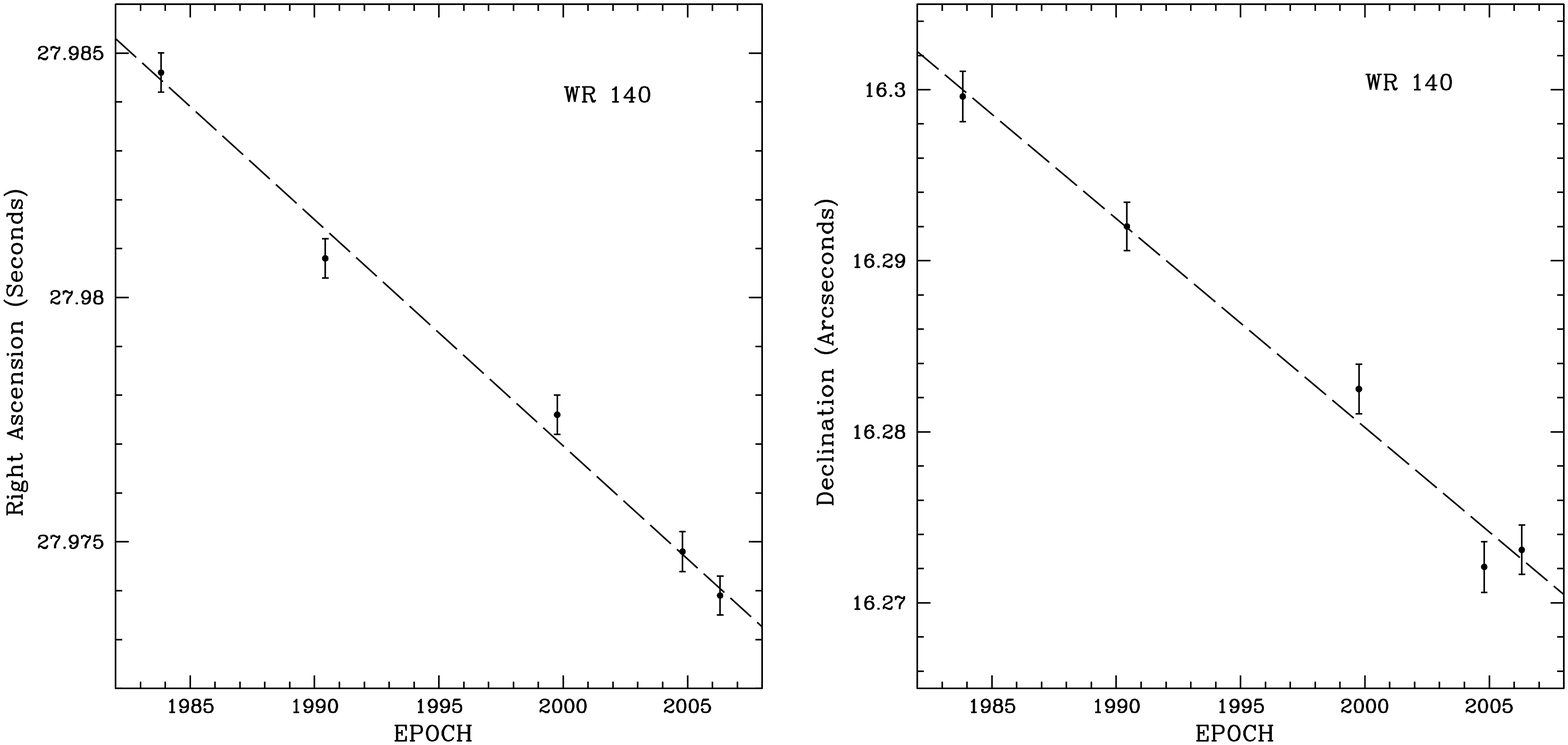}
 \caption{Right ascension (left) and declination (right)
of WR~140 as a function of time. The right ascension is
$20^h~20^m$ and the declination is $43^\circ~51{'}$.
The dashed lines are the least squares
fit to the data.}
  \label{fig1}
\end{figure*}

\subsection{WR 112}

WR~112 belongs to the small group of WC stars with dense dust 
envelopes (van der Hucht et al. 1996).
Assuming that the variable infrared emission is produced by dust
formed in the wind-wind collision zone of a massive binary system,
Marchenko et al. (2002) suggest that this is a WR+OB system with a period of $\sim$25 years.
Assuming a total mass of $\sim$50 $M_\odot$ for the system, this period implies a
semimajor axis of $\sim$30 AU.
The radio study of Chapman et al. (1999) indicates a nonthermal nature for the
centimeter wavelength emission.

In order to measure the proper motions,
we used the epochs given in Table 5.
The positions of the star as a function of time are shown in Figure 2.
The comparison between observed an expected galactic proper
motions indicate that WR 112 has peculiar
motions of order $cos(b) \Delta\mu_l\simeq$ $-16.1\pm5.1$ mas yr$^{-1}$ and $\Delta\mu_b\simeq$ $+3.7\pm3.8$ 
mas yr$^{-1}$, that at a distance of 4.15 kpc is equivalent to peculiar velocities of 
$v_l\simeq -320\pm100~ km~ s^{-1}$ and $v_b\simeq 74\pm76~ km~ s^{-1}$. 
If these large proper motions are confirmed, WR~112 will fall in the category of the runaway stars.

Inspection of Figure 2 suggests that the proper motions of WR~112 in declination
are not smooth. More than a smooth gradient, there seems to be an abrupt change in position 
between epochs 2003.51 and 2004.72. This change is in the order of
$0\rlap.{''}08$, that at a distance of 4.15 kpc is equivalent
to $\sim$330 AU. Since this position shift took
place in a period of $\leq$1.2 years, it implies
(assuming a true physical motion) velocities of
order 1,300 km s$^{-1}$ in the
plane of the sky. This abrupt change cannot be attributed to one of
the members of the binary turning on while the other turned off
because their estimated semimajor axis is only $\sim$30 AU.
Wallace, Moffat, \& Shara (2002) found an optical companion about
$0\rlap.{''}94$ SW of WR~112. This separation is too large to
explain the observed position shift. Finally, we note that
Monnier et al. (2002) have proposed that WR 112 is associated
with a ``pinwheel'' nebula (as it is the case for WR 104 and WR 98a).
These are spiral gaseous structures that surround a few WR stars
and that could be related to the colliding winds found in binary systems
(Monnier et al. 2002). Their presence may produce shifts in
the apparent position of the source. 
In summary, we conclude that the observed position shift in WR 112
should be studied further.


\begin{figure*}
\centering
\includegraphics[scale=0.4, angle=0]{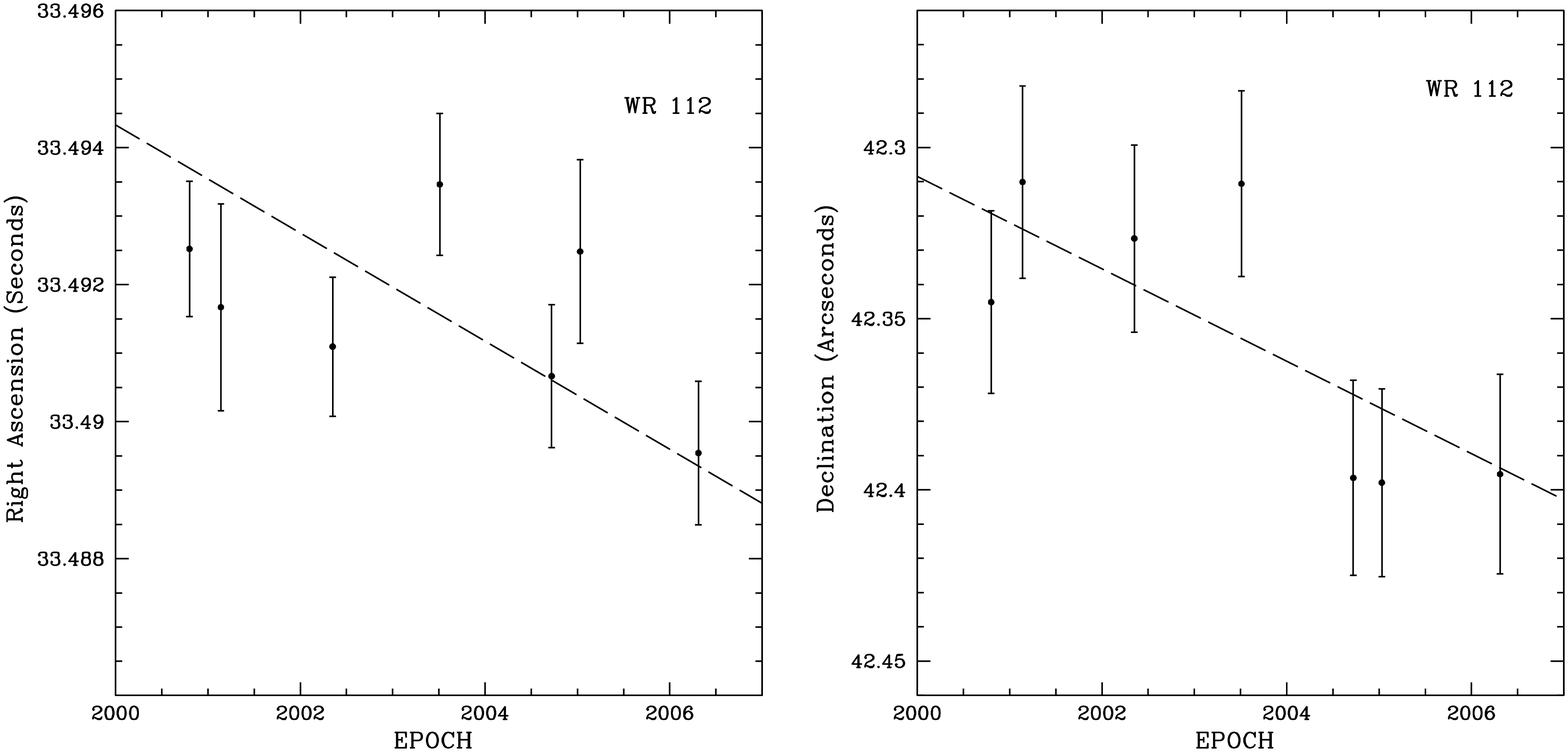}
 \caption{Right ascension (left) and declination (right)
of WR~112 as a function of time. The right ascension is
$18^h~16^m$ and the declination is $-18^\circ~58{'}$. The dashed lines are the least squares
fit to the data.}
  \label{fig2}
\end{figure*}

\begin{table*}[htbp]
\small
  \setlength{\tabnotewidth}{1.0\columnwidth} 
  \tablecols{4} 
  \caption{Archive Data for WR 112}
  \begin{center}
    \begin{tabular}{lccc}\hline\hline
                      &         & Wavelength & Beam  \\
Epoch                 & Project & (cm)& Angular Size $^a$ \\ 
\hline
2000 Oct 17 (2000.80) & AM661	 	& 3.6	& $0\rlap.{''}37 \times 0\rlap.{''}20;~\ -7^\circ$  \\
2001 Feb 21 (2001.14) & AM661 	& 3.6 & $0\rlap.{''}69 \times 0\rlap.{''}49;~  -80^\circ$  \\
2002 May 07 (2002.35) & AM727 	& 3.6 & $0\rlap.{''}36 \times 0\rlap.{''}20;~\ -9^\circ$  \\
2003 Jul 06 (2003.51) & AM766 	& 3.6 & $0\rlap.{''}36 \times 0\rlap.{''}20;~\ -9^\circ$  \\
2004 Sep 21 (2004.72) & AM793 	& 3.6 & $0\rlap.{''}36 \times 0\rlap.{''}20;~\ -9^\circ$  \\
2005 Jan 12 (2005.03) & AM793 	& 3.6 & $0\rlap.{''}38 \times 0\rlap.{''}30;~\ +3^\circ$  \\
2006 Apr 22 (2006.31) & AM831 	& 3.6 & $0\rlap.{''}36 \times 0\rlap.{''}18;~\ -2^\circ$  \\
\hline\hline
\tabnotetext{a}{Major axis $\times$ minor axis; position angle.}
    \label{tab:1}
    \end{tabular}
  \end{center}
\end{table*}

\subsection{WR 125}

WR 125 is believed to be a WC7+O9III binary system  with a period larger
than 18 years (Williams et al. 1994; van der Hucht 2001).

In order to measure the proper motions,
we used the epochs given in Table 6. The positions as a function of time are given in Figure 3.
The observed and expected galactic proper motions
agree well, suggesting that this source
is stationary with respect to its surrounding medium.

\begin{table*}[htbp]
\small
  \setlength{\tabnotewidth}{1.0\columnwidth} 
  \tablecols{4} 
  \caption{Archive Data for WR 125}
  \begin{center}
    \begin{tabular}{lccc}\hline\hline
 & & Wavelength & Beam  \\

Epoch &  Project & (cm) &
Angular Size$^a$ \\
\hline
1985 Feb 16 (1985.13) & AC116 & 6.0
& $0\rlap.{''}42 \times 0\rlap.{''}35;~ -62^\circ$  \\
1993 Jan 16 (1993.04) & AV193 & 6.0
& $0\rlap.{''}36 \times 0\rlap.{''}34;~ -27^\circ$  \\
2000 Oct 19 (2000.80) & AW546 & 6.0
& $0\rlap.{''}37 \times 0\rlap.{''}34;~ 1^\circ$  \\
2002 Mar 03 (2002.17) & AW563 & 6.0
& $0\rlap.{''}39 \times 0\rlap.{''}34;~ -5^\circ$  \\
2004 Dec 06 (2004.93) & AW638 & 6.0
& $1\rlap.{''}11 \times 0\rlap.{''}34;~ 36^\circ$  \\
2006 Feb 18 (2006.13) & AW672 & 6.0
& $0\rlap.{''}39 \times 0\rlap.{''}33;~ -16^\circ$  \\

\hline\hline
\tabnotetext{a}{Major axis $\times$ minor axis; position angle.}
    \label{tab:2}
    \end{tabular}
  \end{center}
\end{table*}

\begin{figure*}
\centering
\includegraphics[scale=0.4, angle=0]{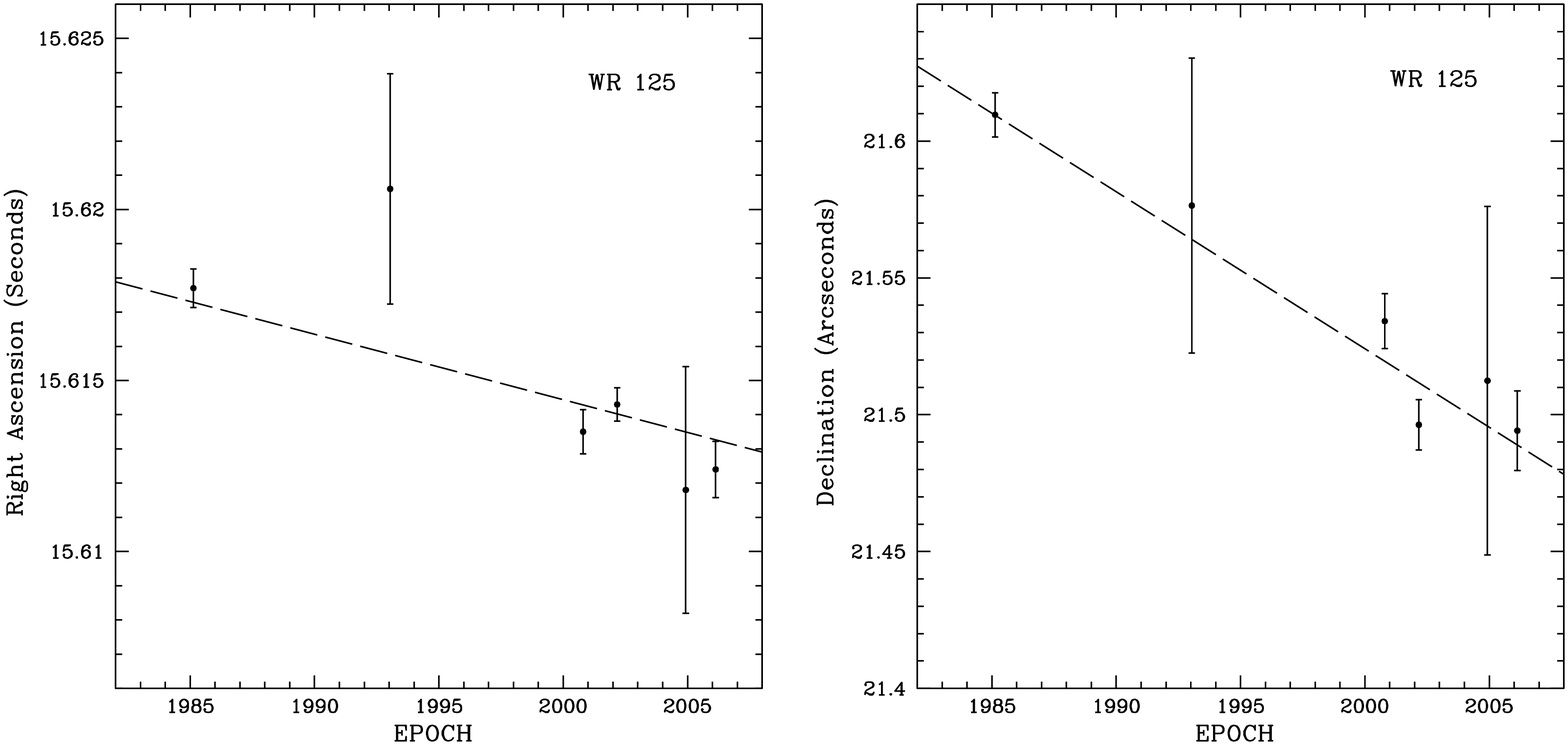}
 \caption{Right ascension (left) and declination (right)
of WR~125 as a function of time. The right ascension is
$19^h~28^m$ and the declination is $19^\circ~33{'}$. The dashed lines are the least squares
fit to the data.}
  \label{fig3}
\end{figure*}

\subsection{WR 145a}

This system is also known as Cyg X-3. It is a WR+C system, where C is for compact 
object, in this case most likely a massive
black hole (Schmutz, Geballe, \& Schild 1996),
although other possibilities are
not discarded (Stark \& Saia 2003). Its period is $\sim$4.8 hours.

The proper motions shown in Figure 4 show a smooth behavior with
the exception of the data point taken in epoch 2000.81, that in declination
shows a significant deviation from the main trend.
This epoch corresponds to the north-south ejection event studied by
Mart\'\i\, Paredes, \& Peracaula (2001). Since this ejection event  
most probably affected the determination of the position, in
particular the declination, this data
point is not included in our fits.

Comparison between the observed and expected galactic
proper motions  given in Table 3, indicates that WR 145a 
has a relative small peculiar motion with respect to its environment, 
$cos(b) \Delta\mu_{l}=-0.7\pm0.3\ mas\ yr^{-1}$ and $\Delta\mu_{b}=-0.3\pm0.2\ mas\ yr^{-1}$. 
Taking the adopted distance of 9 kpc, we have velocities of
$v_{l}=-30\pm 13\ km\ s^{-1}$ and $v_{b}=-13\pm 9\ km\ s^{-1}$, consistent with
this system being approximately stationary with respect to its local standard of rest. 

One of the most important explanation for runaway stars is the so-called 
Blaauw kick: mass unbound suddenly and symmetrically by the supernova 
explosion forces the binary to recoil with a momentum opposite to that of 
the unbound mass at the time of ejection. So, one expects recoil 
velocities in binary systems with a black hole since 
the mechanism by which it is believed that they are formed is a supernova explosion. 
Dhawan et al. (2007) discussed
five binaries with a black hole component, and they showed that two of them 
are not runaways. These authors propose that in the formation of the most massive 
black holes ($\geq10\ M_\odot$), no kick velocities are expected
if they were formed by direct collapse, with no supernova explosion, as proposed
by Fryer \& Kalogera (2001).
Mirabel \& Rodrigues (2003) 
measured the proper motion of Cyg X-1, a $\sim$10 $M_\odot$ black hole,
and found no evidence of peculiar velocities.
The Cyg X-3 system may be another example of this kind of massive black hole formation, 
called by them a ``dark birth''. Another example could be GRS 1915+105 (Dhawan et al. 2007).
These sources may be the ones responsible for the gamma-ray bursts of long duration
in the near Universe without associated luminous supernovae (Della Valle et al. 2006;
Fynbo et al. 2006; Mirabel 2008).

In summary,
the kinematics that we measured for WR 145a appear to indicate 
that it also formed without a significant natal kick. Its small peculiar motion could be
due to a dynamic diffusion process, such as discussed by Dhawan et al. (2007)
for GRS 1915+105.


\begin{table*}[htbp]
\small
  \setlength{\tabnotewidth}{1.0\columnwidth} 
  \tablecols{4} 
  \caption{Archive Data for WR 145a}
  \begin{center}
    \begin{tabular}{lccc}\hline\hline
                      &         & Wavelength & Beam  \\
Epoch                 & Project & (cm)& Angular Size $^a$ \\ 
\hline
1983 Sep 15 (1983.71) & AJ95	 	& 6.0	& $0\rlap.{''}40 \times 0\rlap.{''}36;~ -37^\circ$  \\
1985 Mar 05 (1985.18) & AH172 	& 6.0 & $0\rlap.{''}43 \times 0\rlap.{''}36;~ -69^\circ$  \\
1987 Aug 28 (1987.66) & AC204 	& 6.0 & $0\rlap.{''}24 \times 0\rlap.{''}24;~ -45^\circ$  \\
1997 Jan 10 (1997.03) & AW426 	& 6.0 & $0\rlap.{''}42 \times 0\rlap.{''}36;~ -61^\circ$  \\
2000 Oct 21 (2000.81) & AM669 	& 6.0 & $0\rlap.{''}40 \times 0\rlap.{''}34;~ -23^\circ$  \\
2004 Sep 13 (2004.70) & AR545 	& 3.6 & $0\rlap.{''}22 \times 0\rlap.{''}20;~\ 4^\circ$  \\
2006 May 16 (2006.37) & AM858 	& 3.6 & $0\rlap.{''}26 \times 0\rlap.{''}19;~ -78^\circ$  \\
\hline\hline
\tabnotetext{a}{Major axis $\times$ minor axis; position angle.}
    \label{tab:1}
    \end{tabular}
  \end{center}
\end{table*}

\begin{figure*}
\centering
\includegraphics[scale=0.4, angle=0]{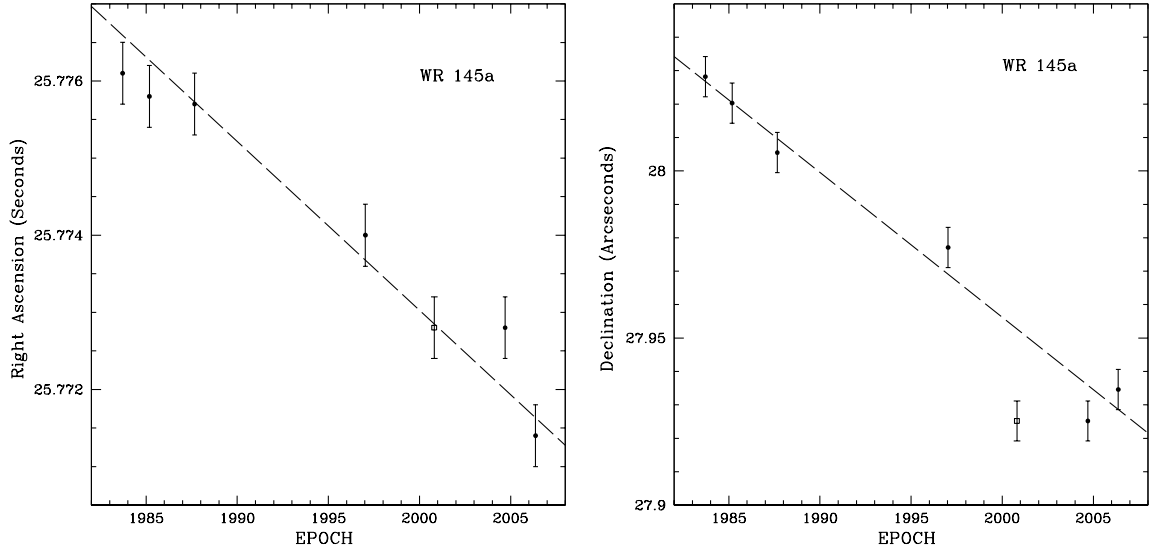}
 \caption{Right ascension (left) and declination (right)
of WR~145a as a function of time. The right ascension is
$20^h~32^m$ and the declination is $40^\circ~57{'}$. The data point for 2000.81
(indicated with an empty square)
was taken during a major ejection event and is not
included in the fit. The dashed lines are the least squares
fit to the data.}
  \label{fig4}
\end{figure*}

\subsection{WR 146}

This is a WR+O system with a 30.8 year period. It is the brightest WR star at 
radio wavelengths. Its companion was first reported by Dougherty et al. (1996) 
and confirmed by Niemela et al. (1998).

In order to measure the galactic proper motion,
we used the epochs given in Table 8. The proper motions are shown in Figure 5.
From the comparison between observed and expected
galactic proper motions given in Table 3, we conclude that the
peculiar motion in $b$ is $\sim16\pm5~ km~ s^{-1}$. This difference is relatively
small and we conclude that
this is not a runaway star. 

\begin{table*}[htbp]
\small
  \setlength{\tabnotewidth}{1.0\columnwidth} 
  \tablecols{4} 
  \caption{Archive Data for WR 146}
  \begin{center}
    \begin{tabular}{lccc}\hline\hline
 & & Wavelength & Beam  \\

Epoch &  Project & (cm) &
Angular Size$^a$ \\
\hline
1991 Jul 12 (1991.53) & AM305 & 3.6
& $0\rlap.{''}22 \times 0\rlap.{''}19;~ -33^\circ$  \\
1996 Oct 26 (1996.82) & AD391 & 6.0
& $0\rlap.{''}48 \times 0\rlap.{''}36;~ -17^\circ$  \\
2004 Oct 1 (2004.75) & AD502  & 3.6
& $0\rlap.{''}22 \times 0\rlap.{''}12;~ 44^\circ$  \\
\hline\hline
\tabnotetext{a}{Major axis $\times$ minor axis; position angle.}
    \label{tab:3}
    \end{tabular}
  \end{center}
\end{table*}

\begin{figure*}
\centering
\includegraphics[scale=0.4, angle=0]{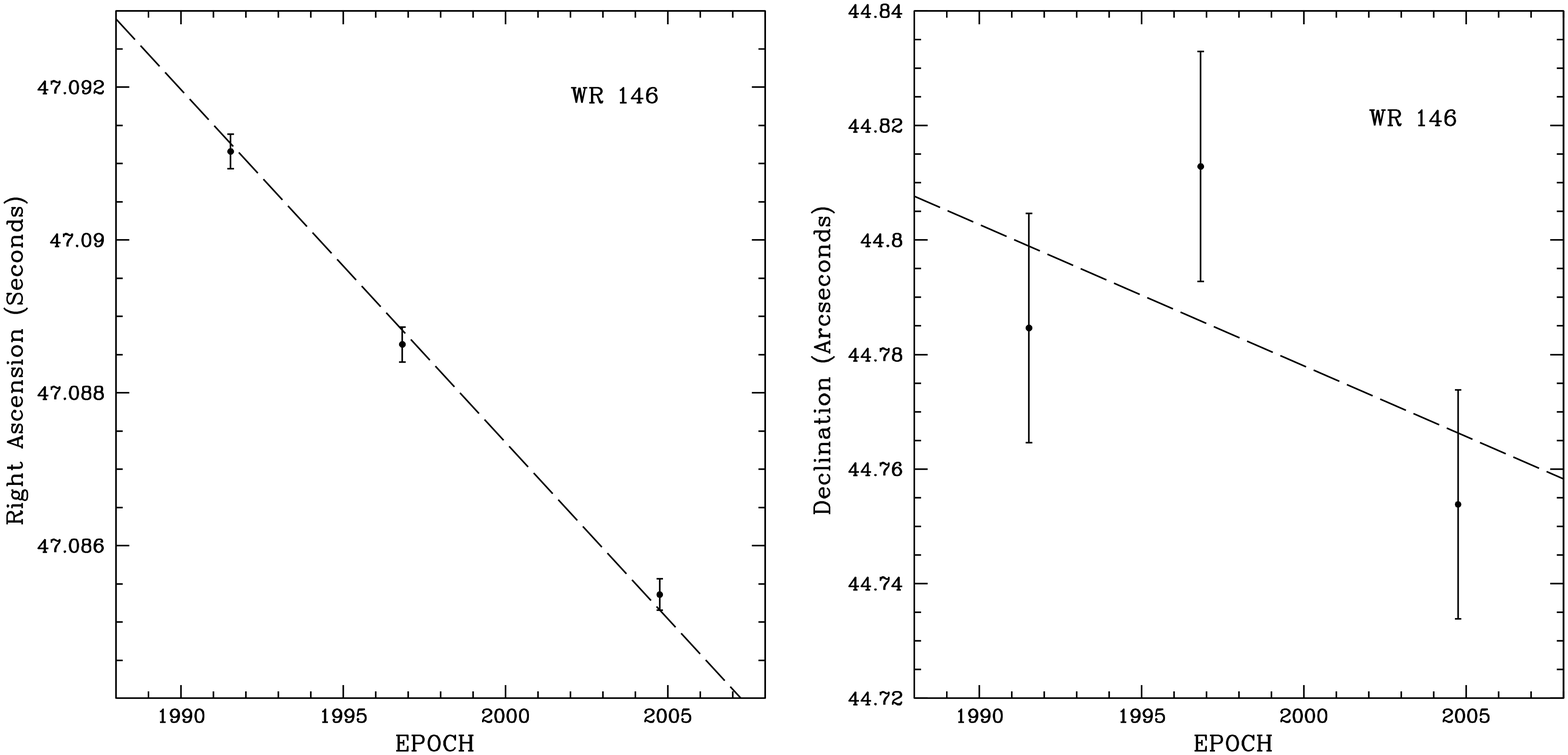}
 \caption{Right ascension (left) and declination (right)
of WR~146 as a function of time. The right ascension is
$20^h~35^m$ and the declination is $41^\circ~22{'}$. The dashed lines are the least squares
fit to the data.}
  \label{fig5}
\end{figure*}

\subsection{WR 147}

This system consists of a WN8 star plus an OB companion 
whose winds are interacting to produce a colliding wind shock. 
The southern component WR 147S is the WN8 star (van der Hucht 2001).
Near-IR images clearly reveal a second source (WR 147N) located
$\approx$0.64 arcsec north of the WN star which was classified as a 
B0.5V star by Williams et al. (1997). An earlier O8-O9 
V-III spectral type was proposed on the basis of Hubble Space Telescope 
(HST) observations (Niemela et al. 1998). The E(B - V) values of WR 
147N and WR 147S are similar (Niemela et al. 1998), confirming that they
are physically associated.

In order to measure the proper motions, 
we used the epochs given in Table 9. The proper motions are shown in Figure 6.
Comparing the observed and expected galactic proper motions, we conclude that WR~147S 
does not have large peculiar proper motions. 

We measured the relative proper motions of the radio source WR 147N with respect to WR 147S,
obtaining
$cos(b) \mu_{l}=-1.6\pm0.7\ mas\ yr^{-1}$ and
$\mu_{b}=-0.4\pm0.6\ mas\ yr^{-1}$. We conclude that these relative proper motions are 
consistent at the 2-$\sigma$ level with no motions (see Figure 7). It is important to emphasize
that the radio source WR 147N is \sl not \rm coincident with the northern star,
but it actually traces the wind interaction zone between the stars
(e. g. Contreras \& Rodr\'\i guez 1999). 

\begin{table*}[htbp]
\small
  \setlength{\tabnotewidth}{1.0\columnwidth} 
  \tablecols{4} 
  \caption{Archive Data for WR 147}
  \begin{center}
    \begin{tabular}{lccc}\hline\hline
 & & Wavelength & Beam  \\

Epoch &  Project & (cm) &
Angular Size$^a$ \\ 
\hline
1985 Feb 16 (1985.13) & AC116 & 6.0
& $0\rlap.{''}37 \times 0\rlap.{''}30;~ -50^\circ$  \\
1995 Jul 21 (1995.55) & AM482 & 3.6
& $0\rlap.{''}22 \times 0\rlap.{''}18;~ -54^\circ$  \\
1996 Dec 14 (1996.95) & AC468 & 3.6
& $0\rlap.{''}20 \times 0\rlap.{''}17;~ 17^\circ$  \\
1999 Sep 03 (1999.67) & AC530 & 6.0
& $0\rlap.{''}43 \times 0\rlap.{''}35;~ -15^\circ$  \\

\hline\hline
\tabnotetext{a}{Major axis $\times$ minor axis; position angle.}
    \label{tab:4}
    \end{tabular}
  \end{center}
\end{table*}

\begin{figure*}
\centering
\includegraphics[scale=0.4, angle=0]{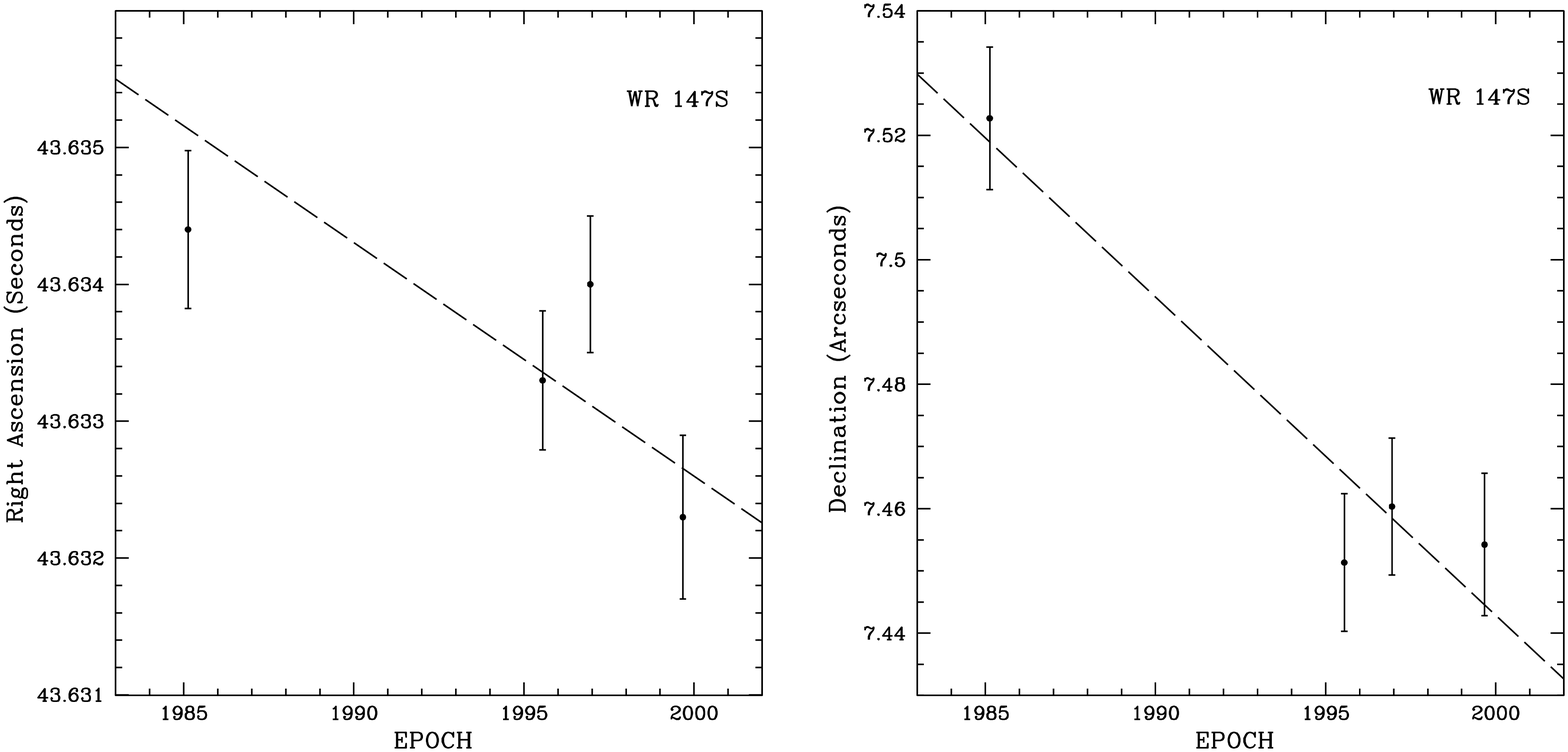}
 \caption{Right ascension (left) and declination (right)
of WR~147S as a function of time. The right ascension is
$20^h~36^m$ and the declination is $40^\circ~21{'}$. The dashed lines are the least squares
fit to the data.}
  \label{fig6}
\end{figure*}

\begin{figure*}
\centering
\includegraphics[scale=0.4, angle=0]{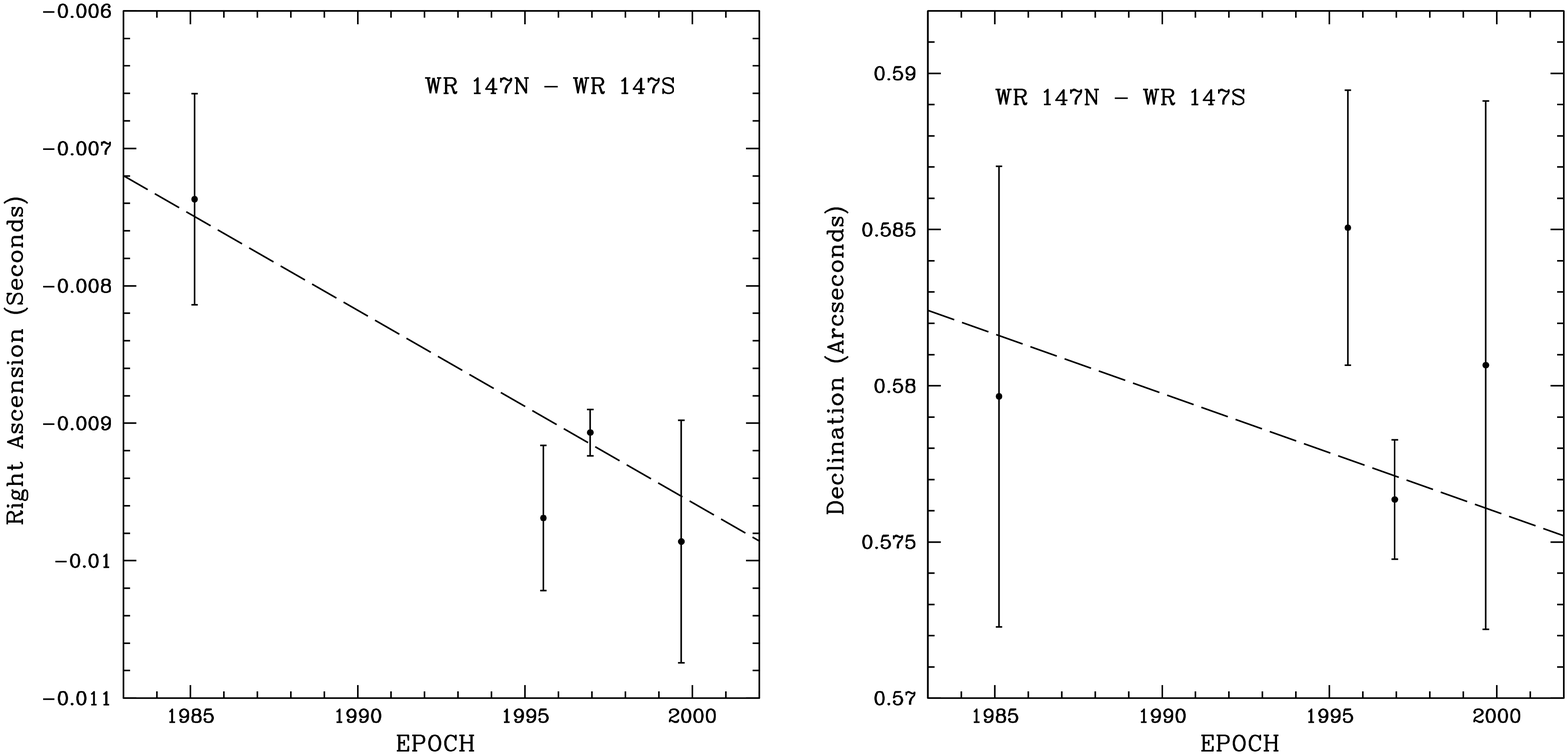}
 \caption{Relative proper motions in right ascension (left) and declination (right)
of WR~147N with respect to
WR~147S as a function of time. 
The dashed lines are the least squares
fit to the data.}
  \label{fig7}
\end{figure*}

\section{Conclusions}

The proper motions measured for WR 125, WR 145a, WR 146 and WR 147 
are in good agreement with the expected proper motions using the galactic 
rotation model of Brand \& Blitz (1993) and the velocity of the Sun 
with respect to the local standard of rest from Dehnen \& Binney (1998).
In the case of WR 145a, the Cyg X-3 binary system, 
proper motions due to the Blaauw kick were expected,
but we do not measure significantly large proper motions with
respect to the local standard of rest of the source.
We propose that this source may be another example
of a dark birth, the collapse of a massive star without a 
supernova ejection.

The proper motions of WR 140 indicate that it has peculiar proper motions
in the order of $\sim30~km~s^{-1}$,
moving away from an original position located in the galactic plane, 
near $l=\ +80\rlap.{^\circ}9$.

The proper motions measured for WR 112 are quite large. This indicates either
we are dealing with a runaway star or that
we may not be measuring the true position of the
star in all the epochs that we used. 

\acknowledgments
We thank I. F. Mirabel for valuable comments. LFR acknowledges the support
of DGAPA, UNAM, and of CONACyT (M\'exico).
This research has made use of the SIMBAD database, 
operated at CDS, Strasbourg, France.


\end{document}